\documentstyle[preprint,aps]{revtex}

\def\rmi{{\rm i}}

\def\rme{{\rm e}}
\def\rmd{{\rm d}}
\def\mbB{{\bbox{B}}}
\def\mbA{{\bbox{A}}}
\def\mbk{{\bbox{k}}}
\def\mbp{{\bbox{p}}}
\def\mbP{{\bbox{P}}}
\def\mbx{{\bbox{x}}}
\def\mbX{{\bbox{X}}}
\def\mbr{{\bbox{r}}}

\begin{document}
\preprint{\leftline{July 2000} \hbox{\hskip -80pt}{UMTG - 233}}
\draft

\tighten

\title{Star operation in Quantum Mechanics
\footnote{This work was supported in part by the National Science
Foundation under grant PHY-9870101}
}   
\author{L. Mezincescu\footnote{e-mail: mezincescu@physics.miami.edu} } 

\address{ Department of Physics, University of Miami, Coral Gables, FL 33124 }

\maketitle

\begin{abstract}
We outline the description of Quantum Mechanics with noncommuting coordinates 
within the framework of star operation. 
We discuss simple cases of integrability.
\end{abstract}

\par
Recently, in connection with developments in string theory \cite{C,S} there 
has been a large interest in the study of noncommutative field theory 
(for a survey of noncommutative field theory and relevant references
see \cite{L}). In noncommutative field theory one deals with fields 
whose base manifold is noncommutative 
(This reminds one of superfield theories where the fermionic coordinates
of the superfields are operators).
It has been shown 
that with the help of the noncommutative, associative star operation the study
of noncommutative field theories can be mapped into that of ordinary 
field theories where ordinary product is replaced by the star product.
In this way generalizations of the Maxwell, Yang-Mills systems and their 
interactions have been produced \cite{C,S}.
\par
Now with respect to the base manifold we can usually formulate a 
Quantum Mechanics, for example for supersymmetric theories we have 
supersymmetric quantum mechanics \cite {W}, not to speak of 
quantum mechanics itself for bosonic coordinates. The fact that the 
noncommutativity of the base manifold can be bypassed with the help of the 
star operation suggests that one can construct a "noncommutative"
quantum mechanics with the same trick.
\par
In this paper we will apply the star operation to Quantum Mechanics.
We will start with the one particle system situation in two dimensions
under the influence of a potential $V({\bf x})$. The Schr\"{o}dinger 
equation is:
\begin{equation}
\rmi \frac{\partial {\Psi ({\mbx},t)}}{\partial t} =  \label{1}
\left[\frac{{{\mbp} }^2}{2m} +V({\mbx})\right]\Psi (\mbx,t)
\end{equation}
and can be obtained from the action:
\begin{equation}
S=\int \rmd t \rmd^2x \bar\Psi \left[ \rmi \frac{\partial}{\partial t}-\frac{{\mbp}^2}
{2m}-V({\mbx})\right]\Psi ({\mbx},t)  \label{2}
\end{equation}
\par
We will formulate this linear Schr\"{o}dinger field theory on a base manifold with 
noncommutative coordinates:
\begin{equation}
\lbrack x'^i,x'^j\rbrack=i\theta ^{ij} \label{3}
\end{equation}  
where $\theta ^{ij}$ is a c-number antisymmetric matrix:
\begin{equation}
\theta ^{ij}=\theta\epsilon^{ij}    \label{4}
\end{equation}
and $\epsilon^{12}=1$ is the totally antisymmetric tensor of rank 2.
\par
As announced, we replace the products above by the star operation:
\begin{equation}
A\star B({\mbx})=e^{\frac{\rmi}{2}\theta ^{ij}\partial^{(1)}_{i}\partial^{(2)}_{j}}A({\mbx _1})B({\mbx _2})\big\vert _{{\mbx _1}={\mbx _2}={\mbx}} \label{5}
\end{equation}
As is well known, under the star operation the terms containing $\frac {\partial}{\partial t}$
and ${\mbp}^2$ are unchanged, however  the potential term will change. 
We are going to have:
\begin{equation}
V(\mbx)\star\Psi(\mbx)=V(\mbx)+\sum_{i=1}^{\infty}\frac{1}{n!}\left (\frac {\rmi}{2}\right )^n
\partial _{i_1}...\partial _{i_n}V(\mbx)\theta ^{i_1j_1}....\theta ^{i_nj_n}
\partial _{j_1}...\partial _{j_n}\Psi(\mbx) \label{6}
\end{equation}
Now, replace $\partial _{jk}$ by $\rmi p_{jk}=\frac{\partial}{\partial {x^{jk}}}$
and introduce:
\begin{equation}
\tilde p _{i_k}=\theta ^{i_kj_k} p_{j_k}   \label{7}
\end{equation}
Take the Fourier transform of $V(\mbx)$, then
\begin{equation}
\partial _{i_1}...\partial _{i_n}V(\mbx)\tilde p_{i_1}...\tilde p_{i_n}\Psi=
\rmi^n\int \rmd^2k \rme^{\rmi\mbk\mbx}V(\mbk)\left (\mbk\tilde\mbp\right )^n \Psi(\mbx) 
\end{equation}
Summing over $n$ in (6) we obtain:
\begin{equation}
V(\mbx)\star\Psi(\mbx)=\int \rmd^2k \rme^{\rmi\mbk\mbx}e^{\frac{\rmi}{2}\tilde\mbp\mbk}V(\mbk)\Psi(\mbx)
\end{equation}
Now using $\mbk\tilde\mbk=0$ and $\left [p_i,x_j\right ]=-\rmi\delta_{ij}$ we obtain:
\begin{equation}
V(\mbx)\star\Psi(\mbx)=V\left (\mbx-\frac{\tilde\mbp}{2}\right )\Psi(\mbx)   \label{10}
\end{equation}
This formula had been inferred before in \cite {D}, which was basically our 
starting point. The star products are evaluated systematically in \cite{Z,T}. 
\par
Introducing the new coordinates $\mbx '$:
\begin{equation}
\mbx '=\mbx-\frac{\tilde\mbp}{2}
\end{equation}
we have (\ref{3}), and the new Hamiltonian is the one in (\ref{1}) with
$\mbx$ replaced by $\mbx '$.
Therefore, from the point of view of star operation, quantum mechanics is being 
redefined by a rearrangement of the phase space variables $\mbx ,\mbp$ into 
$\mbx ' ,\mbp $. It is not hard to realize however that the combination 
$\mbx-\frac{\tilde\mbp}{2}$ is nothing less than the covariant derivative 
in the symmetric gauge 
\begin{equation} 
\mbA =\frac{1}{2}\left ( \mbr\times\mbB \right ) 
\label{11a}
\end{equation}
with proper rescalings, 
corresponding to a constant magnetic field along the $z$-axis. It is therefore 
straightforward to ask whether 
c-noncommutativity of the coordinates, and the presence of 
a magnetic field in the  problem are equivalent. 
We are however not aware of a corresponding $\star$ operation which 
would generate noncommuting coordinates in general gauges, even if 
there exists some freedom in its definition.
Until this question is settled, these are just very intuitive and beautiful 
analogies within this framework of redefining quantum mechanics. 
\par
We will now study a few examples in order to get a clue about the effects 
of star operation.   
It is manifest that for any potential containing inverse powers of 
$\vert\mbx\vert$, the redefinition (\ref{10}) is a horrendous one. Therefore, 
one remains with polynomial interactions, which should be more or less tractable 
and some special situations. The case of an arbitrary potential, which looks 
doable, is the generalization of the Landau electron. We will consider here the 
very asymmetric situation, when one direction is free from interactions, while 
in the other one there is a potential:
\begin{equation}
H=\frac{{{\mbp} }^2}{2m} +V({x})
\end{equation}  
Then, the noncommutative situation is:
\begin{equation}
H=\frac{p_x^2}{2m}+\frac{p_y^2}{2m}+V(x-\theta\frac{p_y}{2})
\end{equation}
and it separates by the choice of wave functions:
\begin{equation}
\Psi=e^{\rmi ky}\Psi (x)
\end{equation}
One gets a well defined one dimensional problem with the potential 
$V(x-\frac{\theta k}{2})$.
\par
We remark that the general case with the potential (\ref{10}) is well suited 
for scattering problems within the perturbative approach, provided they
make sense, and the $-\tilde p/2$ factor generates phases \cite{D} 
when sandwiched between eigenstates of the free Hamiltonian, similar 
to the situation in field theory. In the rest we will restrict to 
quadratic problems which are fully tractable. 
\par
Consider the harmonic oscillator in one dimension. Embedding it in 
a two dimensional noncommutative space:
\begin{equation}
\left [ x,y \right ]=i\theta
\end{equation}
we get by our dynamical principle:
\begin{equation}
H=\frac{p_x^2}{2m}+\frac {k}{2}\left (x-\theta\frac{p_y}{2}\right )^2.
\end{equation}
Setting
\begin{equation}
\frac {k\theta ^2}{4}=\frac{1}{m}
\end{equation}
we get the hamiltonian of a charged particle interacting with a constant 
magnetic field with $qB=2/ \theta$. As before, it should be stressed that
a particular gauge has been chosen for us.
\par
One can also consider an isotropic 2-dimensional oscillator. Then, the
interaction term produces:
\begin{equation}
\frac{k}{2}\left (\mbx-\frac{\tilde\mbp}{2}\right )^2
\label{19}
\end{equation}
which will be related to the symmetric gauge (\ref{11a}). 
The isotropic noncommutative harmonic oscillator (\ref{19}) has been
considered in \cite{J}, in the framework of a study of theories involving 
higher derivatives Chern-Simmons like terms.
\par
Therefore, the purely elastic force is again endowed with a specific 
electromagnetic interaction. The mass of the new system is:
\begin{equation}
\frac{1}{m_R}=\frac{1}{m}+\frac{1}{m'}\;,\;\;m'=\frac{k}{\omega_{\tau}^2}
\;,\;\;\omega_{\tau}=\frac{k\theta}{2}
\end{equation}
with the product $qB$ and renormalized $k$:
\begin{equation}
qB=m_R\omega_{\tau}\;\;\;,\;\;k'=k-m_R\omega_{\tau}^2
\end{equation}
\par
Consider now the two particle system. The wave vector is 
$\Psi (\mbx _1,\mbx _2)$.  Obviously, with two particles we associate the tensor 
product of the two spaces. Therefore we will take $[x_1^i,x_2^j]=0$, otherwise 
it will be more like one particle in a higher dimensional space. 
(That is, before imposing symmetry properties for the wave vector). It is 
natural to take within each set $\mbx_{(i)}$ the same commutation relations, 
even if it is possible to choose them otherwise. Corresponding to $N$ one 
dimensional harmonic oscillators we get a system of $N$ charged particles 
interacting with a constant magnetic field. Replacing $\theta$ with $-\theta$ 
in the commutation relation for some of them, one gets opposite charges.  
Therefore the deformation of a system of such oscillators 
does not produce other forces.
\par
 If however one considers the deformation
of a two particle system interacting through an arbitrary 
$V(\mbx _1-\mbx _2 )$
\begin{equation}
H=\frac{\mbp _1^2}{2m}+\frac{\mbp _1^2}{2m}+ V(\mbx _1-\mbx _2 )  
\end{equation}
the interaction term becomes 
$V\left ( \mbx _1-\mbx _2 -\frac{{\tilde\mbp} _1}{2}  
+\frac{{\tilde\mbp} _2}{2}\right )^2$ (we took identical masses for simplicity). 
Introducing the center of mass and relative coordinates:
\begin{equation}
\mbX =\frac{( \mbx _1+\mbx _2)}{2}\;\;\;\;\;\; 
\bbox{\Delta}=\frac{( \mbx _1-\mbx _2)}{2}
\end{equation}
the Hamiltonian becomes:
\begin{equation}
H=\frac{1}{2(2m)}\mbP ^2+\frac{1}{2(2m)}\mbP ^2_{\Delta}+V\left (2\bbox{\Delta}
-\frac{\tilde\mbP _{\Delta}}{2}\right )
\end{equation}
\par
Therefore, in the new coordinates the motion of the center of mass is free. 
Taking now $V( \mbx _1-\mbx _2)=k/2( \mbx _1-\mbx _2)^2$, we see that the forces 
introduced by the substitution $ \mbx\to\mbx-\tilde\mbp /2$ cannot be only of 
magnetic type in this case, because the center of mass of two charges of 
like sign accelerates when they are in a constant field. This is not the case 
in (24).  Taking now different signs of $\theta$ in the commutation relation of
the corresponding coordinates for the above system, the potential 
$V( \mbx _1-\mbx _2)$ becomes $V(2\bbox{\Delta}-\frac{\tilde\mbP}{2})$ in the 
center of mass frame. In the case of a harmonic force, corresponding to 
the potential $\frac{1}{2}k( \mbx _1-\mbx _2)^2$, the system obtained is like
the one discussed in \cite{D}. The size of the system is proportional to the
momentum of its center of mass. We mention that non-relativistic noncommutative
field theories corresponding to a $\delta$-function interaction between
the particles have been analyzed within the perturbative approach in \cite{G}.
\par
Consider now a quadratic Hamiltonian involving a constant magnetic field, i.e:
\begin{equation}
H=\frac{1}{2m}\left (\mbp-q\mbA\right )^2
\end{equation}
Employing the star operation means going to a $U(1)$ noncommutative 
Maxwell Yang-Mills system with the field strength:
\begin{equation}
F_{\mu\nu}=\partial _{\mu}A_{\nu}-\partial _{\nu}A_{\mu}
-\rmi q (A_{\mu}\star A_{\nu}-A_{\nu}\star A_{\mu})
\end{equation}
and equation of motion,
\begin{equation}
D_{\mu}F_{\mu\nu}=\partial _{\mu}F_{\mu\nu}
-\rmi q(A_{\mu}\star F_{\mu\nu}-F_{\mu\nu}\star A_{\mu})=0,
\end{equation}
and the gauge transformations,
\begin{equation}
\delta\Psi=\rmi q\lambda\star\Psi
\end{equation}
and
\begin{equation}
\delta A_{\mu}=\partial _{\mu}\lambda+\rmi q(\lambda\star A_{\mu}-
A_{\mu}\star\lambda).
\end{equation}
\par
The value of $q$ is fixed. In order to consider the coupling to a "constant
magnetic field" $F_{12}$, one must check whether the symmetric gauge (\ref{11a}) 
satisfies the equation (27). This is indeed the case. However,
\begin{equation}
F_{12}=B\left (1+\frac{q\theta}{4}B\right )
\end{equation}
Therefore, using
\begin{equation}
x_j\star\Psi=\left (x_j-\frac{\tilde p_j}{2}\right )\Psi
\end{equation}
one gets
\begin{equation}
\left (p_i-\frac{qB}{2}\epsilon_{ij}x_j\right )\star\Psi=
\left (1-\frac{qB\theta}{4}\right )\left (p_i-\frac{qB/2}{1-qB\theta/4}\epsilon_{ij}x_j\right )\Psi
\end{equation}
\par
Now, the deformed Hamiltonian is:
\begin{equation}
H=\frac{\left (1-\frac{qB\theta}{4}\right )^2}{2m}
\left (\mbp -\frac{q}{1-\frac{qB\theta}{4}}\mbA\right )^2
\end{equation}
\par
Therefore, the effective couplings and mass change. 
However, the ratio ${{q^2}{B^2}}/m$ remains unchanged. Even if this result 
is obtained in a certain gauge, the magic of the symmetric gauges 
within this approach makes one hope that it may be a true result for the above 
model. We feel that it is certainly worth of further understanding, which 
we lack for the moment.
\par
We have outlined here some of the features of the noncommutative 
Schr\"{o}dinger quantum mechanics, which is just the nonrelativistic one 
particle sector of a quantum field theory. Of course, is not yet clear whether 
the toys presented here have a deeper meaning. However, even at this level the 
connection between the noncommutativity of coordinates and problems containing 
constant magnetic fields seems fascinating. 
\vskip 1cm
\par
The material presented here, come out of discussions with L. Susskind,
to whom I am indebted for encouragement and help.
I also benefitted from a discussion with E. Witten and many discussions with 
G.A. Mezincescu and V. Rittenberg.

\end{document}